\documentclass[aps,prl,twocolumn,superscriptaddress,nofootinbib]{revtex4-1}

\usepackage{amsmath, amssymb}
\usepackage{graphicx}
\usepackage{slashed}

\newcommand{\mgut}{M_{\text{GUT}}}
\newcommand{\mew}{m_{\text{EW}}}
\newcommand{\order}{\mathcal{O}}

\begin{document}

\title{Natural Higgs Mass from Power-Law Running}
\author{Kang-Sin Choi}
\email{kangsin@ewha.ac.kr}
\affiliation{Scranton Honors Program, Ewha Womans University, Seoul 03760, Korea }
\affiliation{Institute of Mathematical Sciences, Ewha Womans University, Seoul 03760, Korea}

\begin{abstract}
The renormalized scalar mass squared is a function of the energy scale and power-runs as its square. Well above the electroweak scale, its dimensionless couplings evolve only slowly, so the Standard Model is approximately scale invariant, and an order-one boundary value supplied at the unification scale is mapped exponentially down to the electroweak scale, much as the QCD scale arises from a dimensionless coupling alone. The 28 orders of magnitude separating the two then measure the smallness of an anomalous dimension, dominated by the top-quark Yukawa coupling, rather than a tuning. Naturalness is thereby recast as the value of an order-one ratio, with no protective symmetry required.
\end{abstract}

\maketitle

The Higgs boson mass parameter $m_h^2$ is close to $(125~\text{GeV})^2$ \cite{ATLAS:Higgs:2012,CMS:Higgs:2012} and 28 orders of magnitude below the scale $\mgut^2 \sim (10^{16}~\text{GeV})^2$ at which the Standard Model is expected to be completed by new physics such as string theory, not necessarily a grand unified theory (GUT). In the Standard Model (SM), the one-loop correction to the scalar mass squared from a field of mass $\mgut$ takes the form $\delta m_h^2 \sim g^2 \mgut^2$, where $g$ is the relevant coupling. The correction overwhelms the observed mass, requiring an apparent cancellation between huge quantities, the bare mass and the loop correction, to one part in $10^{28}$.

This is the hierarchy problem~\cite{Veltman:1981,tHooft:1980,Susskind:1979,Barbieri:Giudice:1988,Giudice:2008}. 't~Hooft gave the symmetry
account~\cite{tHooft:1980}. A mass parameter is protected if setting it to zero increases the symmetry. For fermions, setting the mass to zero,
$m_f \to 0$, restores chiral symmetry; for the scalar,
$m_h \to 0$ restores no symmetry. The resulting corrections
are
\begin{equation}
  \begin{aligned}
    \delta m_f &\propto m_f \log \mgut^2 & &\text{fermion (protected),} \\
    \delta m_h^2 &\propto \mgut^2 & &\text{scalar (unprotected).}
  \end{aligned}
  \label{eq:thooft}
\end{equation}
This was taken to mean that a light scalar is unnatural unless a new symmetry
(supersymmetry~\cite{Dimopoulos:Georgi:1981}, compositeness~\cite{Kaplan:Georgi:1984},
extra dimensions~\cite{ArkaniHamed:1998,Randall:Sundrum:1999}) stabilizes it, or the value is environmentally selected~\cite{Bousso:Polchinski:2000}.

The Large Hadron Collider has found no evidence for such symmetries. We show that none are needed: the fully renormalized mass function exhibits power-law running, the scalar mass squared scaling as the square of the external momentum $q^2$ \cite{Choi:2026yxc,Choi:2023:observables,Choi:2025:Higgs}. This is not an additional assumption but a consequence of standard quantum field theory~\cite{tHooft:1971,tHooft:Veltman:1972}, once the renormalization is carried out completely. The mass function is therefore $\order(\mgut^2)$ at $\mgut^2$ and $\order(m_W^2)$ at $m_W^2$, the expected scaling of an unprotected quantity, and an order-one boundary condition at the unification scale suffices to produce an electroweak-scale Higgs.

The $q^2$-dependence of the scalar self-energy has long been perceived, but only as a part of the divergent, unrenormalized
expression. It appeared in the earliest one-loop
calculations~\cite{Veltman:1981,Fleischer:Jegerlehner:1981};
gauge-invariant formulations were developed in~\cite{Cornwall:1982,Binosi:Papavassiliou:2009}; the physical significance of the quadratic term was explored in~\cite{Jegerlehner:2014,Jegerlehner:2021}; and the
Wilsonian functional renormalization group~\cite{Wilson:Kogut:1974}, which naturally
captures the quadratic scaling, was recently applied to the full SM in~\cite{Garces:2025}. These works treat the $q^2$
dependence within the framework of cutoff-dependent bare parameters or Wilsonian flows. What is new
in~\cite{Choi:2023:observables,Choi:2025:Higgs} is the extraction of a finite, physical mass function in which the $q^2$ dependence is not a cutoff artifact or a source of fine-tuning but the entire observable content of the scalar mass.

\paragraph{The mass function}
On-shell renormalization yields the mass function at every external momentum $q$ \cite{Choi:2026yxc}
\begin{equation}
\begin{split}
  m^2(q^2) &= m_h^2 + \Sigma(q^2)
    - \Sigma(m_h^2)
    - (q^2 - m_h^2) \frac{d\Sigma}{dq^2}(m_h^2) \\
    &\equiv  m_h^2 +\Sigma_{\text{ren}}(q^2) ,
\end{split}
  \label{eq:mren}
\end{equation}
with the dressed propagator $i/(q^2 - m^2(q^2))$ and pole
mass $m_h$ fixed by $m^2(m_h^2) = m_h^2$.\footnote{Above
particle-production thresholds, $\Sigma$ acquires an imaginary
part; $\Sigma_{\text{ren}}$ denotes $\text{Re}\,\Sigma_{\text{ren}}$
throughout.}

The mass function is finite~\cite{Bogoliubov:Parasiuk:1957,Hepp:1966,Zimmermann:1969}
and regularization-independent~\cite{Choi:2023:observables,Choi:2024:decoupling}, so the $q^2$ dependence of $m^2(q^2)$ is a physical effect.
The contribution from gauge bosons can be made gauge-parameter invariant by absorbing non-invariant terms into the vertices, thanks to the Ward--Takahashi identity. The result coincides with the Feynman gauge choice, which we assume from now on \cite{Cornwall:1982,Binosi:Papavassiliou:2009,Papavassiliou:1997fn,Papavassiliou:1997pb,Choi:2026yxc}.  

What runs is the mass {\em function} $m^2(q^2)$, the full momentum-dependent dressing. By construction of eq.~\eqref{eq:mren}, the pole mass is a fixed number; it does not run. The power-law behavior at momenta far from the pole is a prediction; the pole mass itself is data. 
This function $m^2(q^2)$ is the same kind of object as the running coupling
$\alpha_s(q^2)$~\cite{Gross:Wilczek:1973,Politzer:1973}. Both are gauge-invariant, physically measurable and defined at the recoil momentum of the process.

By the on-shell condition $m^2(m_h^2) = m_h^2$, the entire relevant-operator contribution is already taken into account in the measured pole mass: $\Sigma_{\text{ren}}(q^2)$ is the off-shell dressing of this same physical object, not a new $\order(\mgut^2)$ correction beside $m_h^2$. There is nothing to cancel.

\paragraph{Decoupling} The complete renormalization~\eqref{eq:mren} removes the leading sensitivity to heavy fields. A field of mass $M^2 \gg q^2$
contributes~\cite{Choi:2024:decoupling,Choi:2024:stability}
\begin{equation}
  \Sigma_{\text{ren}}^{\text{heavy}}(q^2)
  = \order \left( \frac{(q^2 - m_h^2)^2}{M^2} \right),
  \label{eq:decoupling}
\end{equation}
extending the Appelquist--Carazzone
theorem~\cite{Appelquist:Carazzone:1975} to the super-renormalizable scalar mass. In~\eqref{eq:decoupling}, $m_h$ and $M$ are both pole masses, which are measurable physical quantities, so the suppression is a physical statement, independent of any scheme or convention. 

\begin{figure}[t]
\centering
\includegraphics[width=\columnwidth]{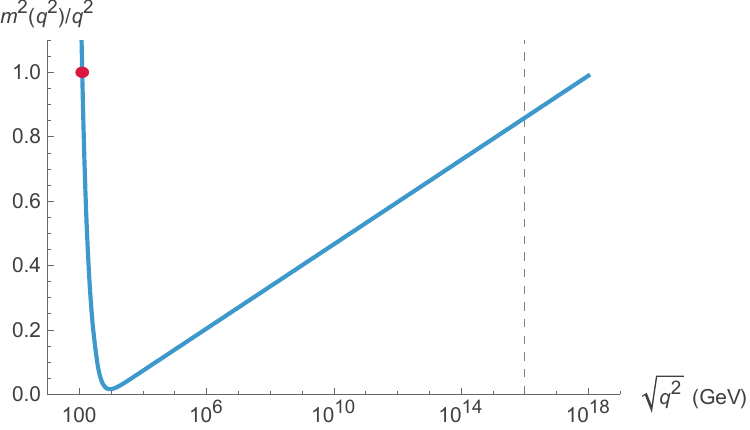}
\caption{The Higgs mass function in units of the external momentum squared,
$m^2(q^2)/q^2$, vs.\ external energy $\sqrt{q^2}$ on a log scale. It shows power-law running up to a $\log q^2$ correction and maintains order-one value up to the GUT scale, which is denoted by the vertical dotted line.
The point corresponds to the measured pole mass. The perturbativity requirement of this function bounds the perturbative domain of the SM at $\sim 10^{18}$~GeV. \label{fig:power}}
\end{figure}

\paragraph{Power-law running of the scalar}

The large-$q^2$ behavior of the renormalized self-energy (see Appendix) is
\begin{equation}
  \begin{aligned}
    \Sigma_{\text{ren}}(\slashed{q})
    &\sim m_f \log q^2 & &\text{fermion,} \\
    \Sigma_{\text{ren}}(q^2)
    &\sim q^2 \log q^2 & &\text{scalar,}
  \end{aligned}
  \label{eq:scaling}
\end{equation}
where the fermion mass function\footnote{The fermion mass function is
$m_f(\slashed{q}) = m_f + \Sigma(\slashed{q}) - \Sigma(m_f) - (\slashed{q}{-}m_f) 
    \left[d\Sigma/d\slashed{q}\right]_{\slashed{q}=m_f}
    \equiv m_f + \Sigma_{\text{ren}}(\slashed{q})$.} satisfies the protected logarithmic running of~\eqref{eq:thooft}.
The crucial difference from the 't~Hooft estimate~\eqref{eq:thooft} is that
the renormalized mass function $m^2(q^2)$ is defined at every scale and is
computed bottom-up from the measured pole mass, whereas the latter gives only a correction at $\mgut$.
Using $m_h = 125.08$~GeV~\cite{ATLAS:2023,CMS:2025}, the complete
one-loop correction to the Higgs mass parameter
(see Appendix) gives, at $\sqrt{q^2} = \mgut = 10^{16}$~GeV,
\begin{equation}
  \frac{m^2(\mgut^2)}{\mgut^2}
  \approx \underbrace{1.1348}_{\text{top}}
    \underbrace{{}-0.1694}_{W}
    \underbrace{{}-0.1086}_{Z}
    +\underbrace{0.0015}_{h}
  = +0.8584 ,
  \label{eq:total_gut}
\end{equation}
which is $\order(1)$.

The total can be written as, in the large $q^2$ limit,
\begin{equation}
\begin{split}  \label{eq:unified}
  \frac{\Sigma_{\text{ren}}^{\text{SM}}}{q^2}
 & \approx \sum_{a=t,W,Z,h} \frac{c_0^a}{16 \pi^2} \log \frac{q^2}{m_a^2} + \text{const} \\
 & \approx \frac{c_0}{16 \pi^2} \log\frac{q^2}{m_h^2} - 0.060 ,
\end{split}
\end{equation}
where 
\begin{equation}\begin{split}
  c_0^t = N_c y_t^2 \approx +2.946, \
  c_0^W =  - g^2 \approx - 0.426,\\
  c_0^Z =  - g_Z^2/2 \approx - 0.274,\
  c_0^h = 3 \lambda \approx +0.387
  \end{split}
\end{equation}
are the one-loop coefficients; threshold
differences are absorbed into the constant. Here, $v = 246.26$~GeV is the Higgs
vacuum expectation value, $N_c = 3$ is the number of quark colors, $y_t$ is the top Yukawa coupling, $g,g_Z$ are weak couplings and $\lambda$ is the quartic coupling.
The leading top coefficient $c_0^t$ is partially canceled by the $W$ and $Z$ contributions, giving the net slope $c_0 \approx 2.24$; the Higgs contribution is negligible at the GUT scale. The last term is the sum of the constant terms from the large-$q^2$ limits (e.g., $-4/3$ from the top in Eq.~\eqref{eq:top_analytic}, plus the $W$ and $Z$ constants).
The coefficient is $\order(10^{-2})$ but multiplies $\log(\mgut^2/\mew^2) \approx 63$, giving an order-one product.

In the $\overline{\text{MS}}$
scheme~\cite{Bardeen:Buras:Duke:Muta:1978} with dimensional
regularization, quadratic powers of the renormalization
scale $\mu^2$ are never generated; they correspond to poles
at $d = 2$, which are invisible at $d = 4 - \epsilon$. The
beta function for $\bar{m}^2(\mu)$ is therefore purely
logarithmic, and the scalar mass {\em appears} to run like
a fermion mass; the power-law $q^2$ dependence,
which lives in the finite parts of the amplitude, goes
unrecognized. $\overline{\text{MS}}$ is not at fault, just a scheme in which the physical scale dependence remains implicit.

The validity of these results requires that the perturbation
theory remain under control at all scales up to
$\mgut$~\cite{Degrassi:2012,Buttazzo:2013}.
The one-loop correction
$\Sigma_{\text{ren}}^{\text{SM}}/q^2 \approx 0.858$ at
$\mgut$ is less than unity, so the propagator denominator
$q^2 - m^2(q^2) = q^2(1 - 0.858) > 0$ remains positive: no
sign change, no ghost, no breakdown. The top quark alone
gives $+1.1348$, which exceeds unity, but the $W$ and
$Z$ contributions ($-0.28$ combined) bring the total below
one; the full SM content is essential (see Appendix). This
partial cancellation is not accidental. Fermion loops carry a
factor $(-1)$ from the closed Grassmann trace
(spin-statistics theorem~\cite{Pauli:1940}), giving the top a positive
contribution while the $W$ and $Z$ contribute with opposite
sign. We have verified that the sign structure persists at
two loops (each fermion loop carries its $(-1)$), so the partial cancellation is a structural property of the SM's
spin content, not a one-loop coincidence. The two-loop correction is suppressed by a further factor of
$\mew^2/(16\pi^2 v^2) \log(\mgut^2/\mew^2)$, giving
$\Sigma^{(2)}/\Sigma^{(1)} \sim 0.1$--$0.2$; this shifts the
total from $0.858$ to roughly $0.9$--$1.0$, which does not
change the order-one character of $m^2(\mgut^2)/\mgut^2$.
That this ratio approaches unity near the high scale is expected; this is the
boundary of the SM's perturbative domain, where the GUT
description takes over.

\paragraph{Constraint on GUT}
\label{sec:focusing}
The scale $\mgut \sim 10^{16}$~GeV indicated by
gauge coupling
unification~\cite{Georgi:Quinn:Weinberg:1974,Langacker:Luo:1991}
is representative of where new physics is expected to appear; in $SU(5)$, the $X$~boson
pole mass provides a more precise candidate for $\mgut$. Below $\mgut$, only SM fields run, and the result~\eqref{eq:total_gut} uses $m_h = 125.08$~GeV as input with no GUT parameters entering.

The ratio $\Sigma_{\text{ren}}^{\text{SM}}/q^2$ grows
logarithmically $\sim \log q^2$ and reaches unity near $10^{18}$~GeV
(Fig.~\ref{fig:power}), an upper bound on the SM's domain of
validity set by the mass function itself. This is independent
of, and consistent with, the standard expectation from gauge
coupling
unification~\cite{Georgi:Quinn:Weinberg:1974,Langacker:Luo:1991},
$\sim 10^{16}$~GeV. The gauge couplings and the Higgs mass
function both point to new physics in the $10^{16}$--$10^{18}$~GeV
window.

The mass function evaluated at the high scale, in units of $\mgut^2$, defines the dimensionless ratio
\begin{equation}
 r  \equiv \frac{m^2(\mgut^2)}{\mgut^2} = \frac{m_h^2
    + \Sigma^{\text{SM}}_{\text{ren}}(\mgut^2)}{ \mgut^2},
  \label{eq:matching}
\end{equation}
which the Standard Model computation~\eqref{eq:total_gut} fixes to $r \approx 0.858$, an order-one number determined entirely by measured parameters. Inverting Eq.~\eqref{eq:matching} for $m_h^2$ gives
\begin{equation}
  m_h^2 = r \mgut^2 
    - \Sigma_{\text{ren}}^{\text{SM}}(\mgut^2) .
  \label{eq:mh_prediction}
\end{equation}
{\em This is the relation traditionally interpreted as fine-tuning:} a small $m_h^2$ on the left against large canceling terms on the right.

Running from the electroweak scale, we have $m_h^2 \ll \mgut^2, \Sigma_{\text{ren}}^{\text{SM}}(\mgut^2)$, so the left-hand side of \eqref{eq:mh_prediction} is negligible. Nevertheless, this equation, with the more explicit form,
\begin{equation} \label{notuning}
  0 \approx \mgut^2  \left[ r
    - \frac{c_0 }{16\pi^2}\log\frac{\mgut^2}{m_h^2} + 0.06 \right],
\end{equation}
relates the electroweak scale $m_h$ and the $\mgut$.
Inverting Eq.~\eqref{notuning} gives
\begin{equation}
  m_h^2  \approx e^{- 16 \pi^2 (r+0.06)/c_0} \mgut^2 .
  \label{eq:exponential}
\end{equation}
{\em There is no fine-tuning.} This is an $\order(1)$ relation in which $m_h^2$ enters only through the logarithm.

\paragraph{Explaining the electroweak scale}

Given GUT parameters, matching equates the GUT boundary condition~\eqref{eq:boundary} to the SM mass function.
We take $SU(5)$ grand unification~\cite{Georgi:Glashow:1974} as a concrete
prototype. The specific parameters differ between UV completions, but the essential structure, namely an order-one boundary condition at the high scale with the SM running below it, does not.
Above $\mgut$, the GUT dynamics determines the Higgs doublet mass at $\mgut$. In $SU(5)$, the boundary condition is
\begin{equation}
  r = \frac{\mu^2 - \frac{3}{2}\lambda_5 V
    + \Sigma^{\text{heavy}}_{\text{GUT}}(\mgut^2)}{\mgut^2} ,
  \label{eq:boundary}
\end{equation}
where $\mu^2$ is the $\mathbf{5}$-Higgs mass parameter, $\lambda_5$
the coupling to the adjoint, $V$ the adjoint VEV, and
$\Sigma^{\text{heavy}}_{\text{GUT}}$ contains loops from
$X$, $Y$ bosons, the triplet, and the adjoint. This defines the matching between the GUT and the SM.

\begin{figure}[t]
\centering
\includegraphics[width=\columnwidth]{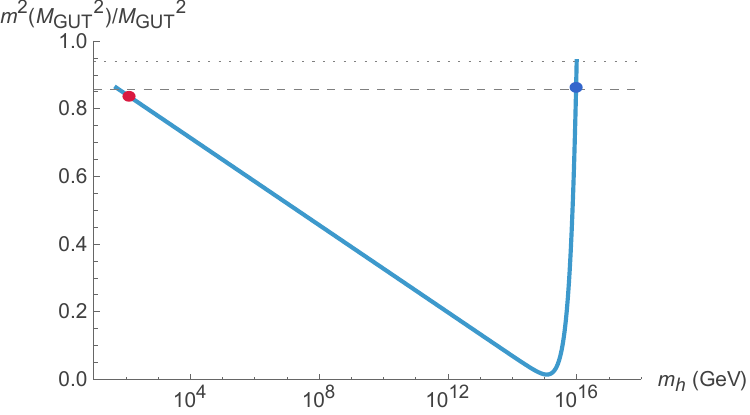}
\caption{The fold of the boundary value, Eq.~\eqref{eq:fold}. The mass function at the GUT scale as a function of the Higgs mass $m_h$. Every value above the minimum at $m_h \approx 0.12\,\mgut$ corresponds to two solutions. The dots mark the two realizations of $r = 0.858$, the quasi-conformal branch with $m_h = 125$~GeV, where the boundary value is the dressing generated by the running, and the heavy branch with $m_h \approx 0.96\,\mgut$, where it is the constant itself. Above $r = 0.94$, its value at $m_h = \mgut$, the heavy solution would require a pole above the matching scale and only the quasi-conformal branch remains. \label{fig:fold}}
\end{figure}

Starting from the UV, the boundary value carries two branches. Writing $x \equiv m_h^2/\mgut^2$ and using that every SM mass inside the logarithm of~\eqref{eq:unified} is proportional to $v$, the boundary value decomposes into a constant part and a dressing,
\begin{equation}
  r(x) = x + \frac{c_0}{16\pi^2}\log\frac{1}{x} - 0.06 ,
  \label{eq:fold}
\end{equation}
and the two terms respond to the electroweak scale in opposite directions: raising $v$ grows the constant linearly but shortens the logarithm. The function therefore has a minimum at $x = c_0/16\pi^2 \approx 0.014$, that is, at $m_h \approx 0.12\,\mgut$, and every boundary value above $r_{\min} \approx 0.015$ is realized by exactly two solutions (Fig.~\ref{fig:fold}).

On the light branch the constant is negligible and the boundary value is almost entirely the dressing accumulated by the running: for $m_h = 125$~GeV, $r = 0.858 = 1.6\times 10^{-28} + 0.858$. This is the quasi-conformal branch, on which the inversion~\eqref{eq:exponential} holds. In the great desert, where $\mew \ll \sqrt{q^2} \ll \mgut$, the dimensionless couplings of the SM run only slowly and no new mass thresholds intervene, so the theory is approximately scale invariant~\cite{Bardeen:1995kv} and the mass function scales canonically, $m^2(q^2) \propto q^2$, up to the slow logarithm whose slope $c_0/16\pi^2$ is an anomalous-dimension coefficient. The 28 orders of magnitude between $m_h^2$ and $\mgut^2$ then measure the smallness of this anomalous dimension, not a tuning, with logarithmic sensitivity $d\log x/dr = -16\pi^2/c_0$ comparable to that of $\Lambda_{\text{QCD}}$ on the unification-scale gauge coupling. The mechanism resembles dimensional transmutation: $\Lambda_{\text{QCD}}$ is generated from a dimensionless coupling alone, whereas here an order-one ratio $r$ given at $\mgut$ is mapped exponentially downward, the slow logarithm playing the role of the running coupling. This anomalous dimension is dominated by the top-quark Yukawa coupling $y_t$, with sizable contributions from $W$, $Z$ and Higgs bosons.

On the heavy branch the boundary value is almost entirely the constant. The same $r = 0.858$ is realized by a doublet of pole mass $\approx 0.96\,\mgut$, whose mass function is nearly flat in $q^2$ and whose vacuum leaves the electroweak symmetry unbroken at low energies.\footnote{On the heavy branch all SM masses scale with the heavy vacuum, so Eq.~\eqref{eq:fold}, derived in the large-$q^2$ limit with the broken-phase spectrum, is schematic near $x \sim 1$; the exact one-loop evaluation confirms the second solution.} This heavy branch is off-critical, the constant dominating and scaling broken at every scale below $\mgut$, and it inverts linearly, $m_h^2 \approx r\,\mgut^2$.

The fold is itself a consequence of the power-law running and explains why the scalar is special among running quantities. A gauge coupling at $\mgut$ is a pure logarithm of the infrared data, and the fermion mass function is protected by~\eqref{eq:scaling}, its dressing proportional to $m_f^2$; both are single-valued. The scalar is the unique unprotected case. Its dressing grows as $q^2 \log q^2$ and overtakes any light constant, folding the boundary value.

The constraint on the GUT~\eqref{eq:boundary} is the order-one range $0 < r \le 1$, as long as the GUT scale is below $\sim 10^{18}$~GeV. The fold's two solutions raise the question of which branch is realized. For matching above $\sim 2\times 10^{17}$~GeV, the string scale, the boundary value exceeds $r(x{=}1) = 0.94$ and only the quasi-conformal branch survives, so the boundary value alone selects the electroweak-scale solution, Eq.~\eqref{eq:exponential}.

\paragraph{From naturalness to flavor}
With nothing to cancel, the GUT delivers a value $0<r\le 1$, and the residual question is the specific value of an order-one ratio, not a quadratic sensitivity requiring new symmetries at the TeV scale.
It is a question about the specific value of an order-one coupling, not protected by
any symmetry, yet never regarded as unnatural. Even the
electron Yukawa $y_e \sim 10^{-6}$, six orders of magnitude
below the top Yukawa $y_t \approx 1$ and far more
``unexplained'' than any $\order(1)$
ratio, is not considered a naturalness crisis. The specificity of the GUT-scale ratio
belongs to the same category of questions about coupling
values, approachable through whatever ``flavor'' symmetry organizes the couplings.

The absence of protection is not the disease. It is what makes the scalar power-run, recasting the hierarchy. The large ratio measures the smallness of an anomalous dimension, leaving only the value of an order-one boundary condition.
An experimental test through off-shell Higgs
production~\cite{Kauer:Passarino:2012,Caola:Melnikov:2013}
at the LHC is discussed in
Ref.~\cite{Choi:2025:Higgs}.

As a more speculative outlook, the same argument may apply to the cosmological constant $\Lambda$ with mass dimension four. In flat space, it is a constant with no external momentum. In curved space, it couples to the graviton via $\sqrt{\det g}\,\Lambda$, through which $\Lambda(q^2)$ acquires external momentum dependence carried by the graviton. Power-law running would then give $\Lambda(q^2)/q^4 = \order(1)$ at every scale $q^2$. We call for the explicit construction of $\Lambda(q^2)$, which requires a renormalizable formulation of quantum gravity.

\begin{acknowledgments}
The author is grateful to Jaehoon Jeong, Hyun-Min Lee and Matthew McCullough for discussions. He thanks especially Taehyun Jung for valuable comments. This work is partly supported by the grant RS-2023-00277184 of the National Research Foundation of Korea. The author used Claude (Anthropic) for editing and discussion during the manuscript preparation. The final content was reviewed and approved by the author, who takes full responsibility for the work.
\end{acknowledgments}

\appendix
\section{Explicit one-loop formulas}
\label{app:formulas}

We collect the renormalized one-loop self-energies from
Ref.~\cite{Choi:2025:Higgs} that yield the
result~\eqref{eq:total_gut}. The total mass function is
$m^2(q^2) = m_h^2 +
\Sigma_{\text{ren}}^{\text{SM}}(q^2)$, where
$\Sigma_{\text{ren}}^{\text{SM}} = \sum_{a=t,W,Z,h}
\text{Re} \Sigma_{1,\text{ren}}^a(q^2)$, where $x$ is the
Feynman parameter and $m$ is the mass of the particle in the
loop.

\textit{Top quark:}
\begin{multline}
  \Sigma_{1,\text{ren}}^t(q^2)
  = \frac{3m_t^2 N_c}{4\pi^2 v^2}
    \int_0^1 \!\! dx \biggl[
      \Delta_{m_t}(q^2)
      \log\frac{\Delta_{m_t}(m_h^2)}{\Delta_{m_t}(q^2)} \\
      - (q^2 - m_h^2) x(1{-}x)
    \biggr],
  \label{eq:sigma_f}
\end{multline}
with $\Delta_m(q^2) \equiv m^2 - x(1{-}x) q^2$.

\textit{$W$ boson} (transverse + Goldstone + ghost):
\begin{multline}
  \Sigma_{1,\text{ren}}^W(q^2)
  = \frac{-1}{16\pi^2 v^2}
    \int_0^1 \!\! dx \biggl[
      L_{m_W}(q^2)
      \log\frac{\Delta_{m_W}(m_h^2)}{\Delta_{m_W}(q^2)} \\
      - (q^2 - m_h^2) 
        \frac{x(1{-}x) L_{m_W}(m_h^2)}{\Delta_{m_W}(m_h^2)}
    \biggr],
  \label{eq:sigma_W}
\end{multline}
with $L_M(q^2) \equiv m_h^4 - 2q^2(1{+}3x^2)M^2 + 10M^4$.

\textit{$Z$ boson:}  Same structure with an overall factor
$1/2$ and $m_W \to m_Z$.

\textit{Higgs self-coupling:}
\begin{multline}
  \Sigma_{1,\text{ren}}^h(q^2)
  = \frac{-9m_h^4}{32\pi^2 v^2}
    \int_0^1 \!\! dx \biggl[
      \log\frac{\Delta_{m_h}(m_h^2)}{\Delta_{m_h}(q^2)} \\
      - (q^2 - m_h^2) 
        \frac{x(1{-}x)}{\Delta_{m_h}(m_h^2)}
    \biggr].
  \label{eq:sigma_h}
\end{multline}

At $q^2 \gg m_t^2$, the top contribution simplifies to
\begin{equation}
  \frac{\Sigma_{1,\text{ren}}^t}{q^2}
  \approx \frac{N_c m_t^2}{4\pi^2 v^2}
    \biggl[\frac{1}{2}\log\frac{q^2}{m_t^2}
    - \frac{4}{3}\biggr],
  \label{eq:top_analytic}
\end{equation}
exhibiting the characteristic $q^2 \log q^2$ power-law
behavior. The self-energy grows as $q^2$ times a logarithm,
so $\Sigma/q^2$ grows logarithmically, but the mass function
$m^2(q^2) \approx m_h^2 + \Sigma_{\text{ren}}(q^2)$ grows as
$q^2$ itself. The $W$ and $Z$ contributions have the same
power-law structure with opposite sign,
\begin{equation}
  \frac{\Sigma_{1,\text{ren}}^W}{q^2}
  \approx -\frac{m_W^2}{4\pi^2 v^2}
     \log\frac{q^2}{m_W^2} + \text{const} ,
  \label{eq:W_analytic}
\end{equation}
and similarly for $Z$ with a factor $1/2$.
The numerical evaluation of these integrals at
$\sqrt{q^2} = 10^{16}$~GeV, using $m_t = 172.56$~GeV,
$m_W = 80.37$~GeV, $m_Z = 91.19$~GeV,
$v = 246.26$~GeV~\cite{PDG:2024},
yields the sector-by-sector values quoted in
Eq.~\eqref{eq:total_gut}.


\begin{thebibliography}{99}

\bibitem{ATLAS:Higgs:2012}
  ATLAS Collaboration,
  Phys.\ Lett.\ B \textbf{716}, 1 (2012).

\bibitem{CMS:Higgs:2012}
  CMS Collaboration,
  Phys.\ Lett.\ B \textbf{716}, 30 (2012).

\bibitem{Veltman:1981}
  M.~J.~G.~Veltman,
  Acta Phys.\ Polon.\ B \textbf{12}, 437 (1981).

\bibitem{tHooft:1980}
  G.~'t~Hooft,
  NATO ASI Ser.\ B \textbf{59}, 135 (1980).

\bibitem{Susskind:1979}
  L.~Susskind,
  Phys.\ Rev.\ D \textbf{20}, 2619 (1979).

\bibitem{Barbieri:Giudice:1988}
  R.~Barbieri and G.~F.~Giudice,
  Nucl.\ Phys.\ B \textbf{306}, 63 (1988).

\bibitem{Giudice:2008}
  G.~F.~Giudice,
  ``Naturally Speaking: The Naturalness Criterion
  and Physics at the LHC,''
  arXiv:0801.2562 [hep-ph].

\bibitem{Dimopoulos:Georgi:1981}
  S.~Dimopoulos and H.~Georgi,
  Nucl.\ Phys.\ B \textbf{193}, 150 (1981).

\bibitem{Kaplan:Georgi:1984}
  D.~B.~Kaplan and H.~Georgi,
  Phys.\ Lett.\ B \textbf{136}, 183 (1984).

\bibitem{ArkaniHamed:1998}
  N.~Arkani-Hamed, S.~Dimopoulos and G.~Dvali,
  Phys.\ Lett.\ B \textbf{429}, 263 (1998).

\bibitem{Randall:Sundrum:1999}
  L.~Randall and R.~Sundrum,
  Phys.\ Rev.\ Lett.\ \textbf{83}, 3370 (1999).

\bibitem{Bousso:Polchinski:2000}
  R.~Bousso and J.~Polchinski,
  JHEP \textbf{06}, 006 (2000).

%\cite{Choi:2026yxc}
\bibitem{Choi:2026yxc}
K.-S.~Choi,
``The Gauge-Invariant Mass Function,''
[arXiv:2604.04569 [hep-th]].

\bibitem{Choi:2023:observables}
  K.-S.~Choi,
  ``On the observables of renormalizable interactions,''
  J.\ Korean Phys.\ Soc.\ \textbf{84}, 591 (2024),
  arXiv:2310.00586 [hep-ph].

\bibitem{Choi:2025:Higgs}
  K.-S.~Choi,
  ``One-Loop Correction to the Higgs Mass,''
  arXiv:2506.18667 [hep-ph].

\bibitem{tHooft:1971}
  G.~'t~Hooft,
  Nucl.\ Phys.\ B \textbf{35}, 167 (1971).

\bibitem{tHooft:Veltman:1972}
  G.~'t~Hooft and M.~J.~G.~Veltman,
  Nucl.\ Phys.\ B \textbf{44}, 189 (1972).

\bibitem{Fleischer:Jegerlehner:1981}
  J.~Fleischer and F.~Jegerlehner,
  Phys.\ Rev.\ D \textbf{23}, 2001 (1981).

\bibitem{Cornwall:1982}
  J.~M.~Cornwall,
  Phys.\ Rev.\ D \textbf{26}, 1453 (1982).

\bibitem{Binosi:Papavassiliou:2009}
  D.~Binosi and J.~Papavassiliou,
  Phys.\ Rept.\ \textbf{479}, 1 (2009).

\bibitem{Jegerlehner:2014}
  F.~Jegerlehner,
  arXiv:1406.3658 [hep-ph].

\bibitem{Jegerlehner:2021}
  F.~Jegerlehner,
  arXiv:2106.00862 [hep-ph].

\bibitem{Wilson:Kogut:1974}
  K.~G.~Wilson and J.~Kogut,
  Phys.\ Rept.\ \textbf{12}, 75 (1974).

\bibitem{Garces:2025}
  J.~P.~Garc\'es, F.~Goertz and \'A.~Pastor~Guti\'errez,
  JHEP \textbf{10}, 134 (2025),
  arXiv:2506.15919 [hep-ph].


  \bibitem{Bogoliubov:Parasiuk:1957}
  N.~N.~Bogoliubov and O.~S.~Parasiuk,
  Acta Math.\ \textbf{97}, 227 (1957).

\bibitem{Hepp:1966}
  K.~Hepp,
  Comm.\ Math.\ Phys.\ \textbf{2}, 301 (1966).


\bibitem{Zimmermann:1969}
  W.~Zimmermann,
  Comm.\ Math.\ Phys.\ \textbf{15}, 208 (1969).

\bibitem{Choi:2024:decoupling}
  K.-S.~Choi,
  ``Renormalization, Decoupling and the Hierarchy Problem,''
  arXiv:2408.06406 [hep-ph].
  
%\cite{Papavassiliou:1997fn}
\bibitem{Papavassiliou:1997fn}
J.~Papavassiliou and A.~Pilaftsis,
``Effective charge of the Higgs boson,''
Phys.\ Rev.\ Lett.\ \textbf{80}, 2785 (1998).

%\cite{Papavassiliou:1997pb}
\bibitem{Papavassiliou:1997pb}
J.~Papavassiliou and A.~Pilaftsis,
%``Gauge and renormalization group invariant formulation of the Higgs boson resonance,''
Phys.\ Rev.\ D \textbf{58}, 053002 (1998).

\bibitem{Bardeen:Buras:Duke:Muta:1978}
  W.~A.~Bardeen, A.~J.~Buras, D.~W.~Duke
  and T.~Muta,
  Phys.\ Rev.\ D \textbf{18}, 3998 (1978).

\bibitem{Gross:Wilczek:1973}
  D.~J.~Gross and F.~Wilczek,
  Phys.\ Rev.\ Lett.\ \textbf{30}, 1343 (1973).

\bibitem{Politzer:1973}
  H.~D.~Politzer,
  Phys.\ Rev.\ Lett.\ \textbf{30}, 1346 (1973).

\bibitem{Choi:2024:stability}
  K.-S.~Choi,
  ``Stability of the Scalar Mass against Loop Corrections,''
  arXiv:2410.21118 [hep-ph].

\bibitem{Appelquist:Carazzone:1975}
  T.~Appelquist and J.~Carazzone,
  Phys.\ Rev.\ D \textbf{11}, 2856 (1975).

\bibitem{ATLAS:2023}
  ATLAS Collaboration,
  Phys.\ Lett.\ B \textbf{847}, 138315 (2023).

\bibitem{CMS:2025}
  CMS Collaboration,
  Phys.\ Rev.\ D \textbf{111}, 092014 (2025).

\bibitem{Georgi:Quinn:Weinberg:1974}
  H.~Georgi, H.~R.~Quinn and S.~Weinberg,
  Phys.\ Rev.\ Lett.\ \textbf{33}, 451 (1974).

\bibitem{Langacker:Luo:1991}
  P.~Langacker and M.~Luo,
  Phys.\ Rev.\ D \textbf{44}, 817 (1991).

\bibitem{Georgi:Glashow:1974}
  H.~Georgi and S.~L.~Glashow,
  Phys.\ Rev.\ Lett.\ \textbf{32}, 438 (1974).

\bibitem{Degrassi:2012}
  G.~Degrassi \textit{et al.},
  JHEP \textbf{08}, 098 (2012).

\bibitem{Buttazzo:2013}
  D.~Buttazzo \textit{et al.},
  JHEP \textbf{12}, 089 (2013).

\bibitem{Pauli:1940}
  W.~Pauli,
  Phys.\ Rev.\ \textbf{58}, 716 (1940).

\bibitem{Kauer:Passarino:2012}
  N.~Kauer and G.~Passarino,
  JHEP \textbf{08}, 116 (2012).

\bibitem{Caola:Melnikov:2013}
  F.~Caola and K.~Melnikov,
  Phys.\ Rev.\ D \textbf{88}, 054024 (2013).

\bibitem{PDG:2024}
  S.~Navas \textit{et al.} (Particle Data Group),
  Phys.\ Rev.\ D \textbf{110}, 030001 (2024).

\bibitem{Bardeen:1995kv}
  W.~A.~Bardeen,
  ``On naturalness in the standard model,''
  FERMILAB-CONF-95-391-T (1995).

\end{thebibliography}
\end{document}